\documentclass[11pt,a4paper]{article}

\usepackage{jcappub}
\usepackage{multicol}
\usepackage{indentfirst}

\usepackage[UKenglish]{babel}
\usepackage[utf8]{inputenc}
\usepackage[T1]{fontenc}
\usepackage{amsfonts}
\usepackage{verbatim}
\usepackage{hyphenat}

\usepackage{graphicx}
\graphicspath{ {./Figures/} }
\def\WidthScale{0.7}
\usepackage{float}
\usepackage{epsfig}
\usepackage{rotate}
\usepackage{subcaption}
\usepackage{booktabs}
\usepackage{tablefootnote}
\usepackage{afterpage}

\usepackage{amsmath}
\usepackage{amssymb}
\usepackage{mathrsfs}
\usepackage{empheq}
\usepackage{relsize}
\usepackage{bm}
\allowdisplaybreaks

\usepackage{physics}
\usepackage{tensor}

\usepackage{cleveref}
\usepackage{orcidlink}

\usepackage{xcolor}
\usepackage[colorinlistoftodos, tickmarkheight=0.2cm, textsize=tiny]{todonotes}
  
\usepackage{enumitem}
\numberwithin{equation}{section}
\numberwithin{figure}{section}
\numberwithin{table}{section}

\newcommand{\bs}[1]{\boldsymbol{#1}}

\renewcommand{\H}{\mathcal{H}}

\renewcommand{\P}{\mathcal{P}}

\DeclareMathAlphabet\mathbfcal{OMS}{cmsy}{b}{n}
\newcommand{\A}{\mathbfcal{A}}

\newcommand{\biposh}{BipoSH}

\newcommand{\largesum}{\mathop{\mathlarger{\sum}}}

\title{Probing the Cosmological Principle\\ with CMB lensing and cosmic shear}

\author[a]{James Adam\orcidlink{0000-0002-2585-0488},}
\author[a,b]{Roy Maartens\orcidlink{0000-0001-9050-5894},}
\author[c,a]{Julien Larena\orcidlink{0000-0003-3301-0435},}
\author[d,a]{\\ Chris Clarkson\orcidlink{0000-0001-7363-0722},}
\author[e]{Ruth Durrer\orcidlink{0000-0001-9833-2086}}

\affiliation[a]{Department of Physics \& Astronomy, University of the Western Cape, Cape Town 7535, South Africa}
\affiliation[b]{National Institute for Theoretical \& Computational Sciences, 
Cape Town 7535, South Africa}
\affiliation[c]{Laboratoire Univers et Particules de Montpellier, CNRS \& Université de Montpellier,\\ Montpellier 34090, France}
 \affiliation[d]{Department of Physics \& Astronomy, Queen Mary University of London,\\ London E1 4NS, United Kingdom}
 \affiliation[e]{D\'epartement de Physique Th\'eorique, Universit\'e de Gen\`eve, CH-1211 Gen\`eve 4, Switzerland}

\emailAdd{james.g.adam1997@gmail.com} 

\abstract{The standard cosmological model assumes the Cosmological Principle. However, recent observations hint at possible violations of isotropy on large scales, possibly through late-time anisotropic expansion. Here we investigate the potential of cross-correlations between CMB lensing convergence $\kappa$ and galaxy cosmic shear $B$-modes as a novel probe of such late-time anisotropies. Our signal-to-noise forecasts reveal that information from the $\kappa$-$B$ cross-correlation is primarily contained on large angular scales ($\ell \lesssim 200$). We find that this cross-correlation for a Euclid-like galaxy survey is sensitive to anisotropy at the percent level. Making use of tomography yields a modest improvement of $\sim 20\%$ in detection power. Incorporating the galaxy $E$-$B$ cross-correlations would further enhance these constraints.}

\begin{document}

\maketitle


\section{Introduction}

The standard model of cosmology rests on the Cosmological Principle, which asserts that the Universe is both homogeneous and isotropic on sufficiently large scales. Developing novel and complementary methods to probe possible violations of these assumptions is a major focus of contemporary research \cite{Euclid:2022ucc, CosmoVerse:2025alb} (see also \cite{Clarkson:2010uz,Maartens:2011yx}). It is much more difficult to test homogeneity than isotropy and thus it is typical to assume the Copernican Principle --- that we are not at a special location in the Universe. Then any violation of large-scale isotropy implies a violation of the Cosmological Principle.
Observations of the cosmic microwave background (CMB) \cite{Stoeger:1994qs,martinez-gonzalez_delta_1995, maartens_anisotropy_1996,bunn_how_1996, kogut_limits_1997, saadeh_how_2016}, baryon acoustic oscillations \cite{akarsu_testing_2023}, and big bang nucleosynthesis \cite{rothman_effects_1982, rothman_nucleosynthesis_1984, matzner_conjecture_1986, campanelli_helium-4_2011}, together with theoretical considerations of isotropisation during inflation \cite{anninos_how_1991, pitrou_predictions_2008}, generally disfavour large-scale anisotropy in the primordial and early Universe (see \cite{aluri_is_2023} for an alternate perspective). However, it is conceivable that effects from structure formation or in the physics of the dark sector could generate late-time anisotropy in the Universe (see e.g. \cite{palacios-cordoba_anisotropic_2025} and the review \cite{CosmoVerse:2025alb}).

Significant anisotropy in the late-time universe would imprint a signature in the large-scale structure --- for example, via a preferred direction from an anomalous bulk flow \cite{CosmoVerse:2025alb}.
The kinematic dipole in galaxy surveys due to the observer's motion should agree with the CMB kinematic dipole according to the Cosmological Principle. However, measurements indicate that, while these dipoles are consistent in direction, the velocities may be significantly different (see \cite{CosmoVerse:2025alb}). Similarly, there are indications that the bulk flow of matter has an anomalously large magnitude. These results hint at a possible violation of isotropy via a preferred direction. Anisotropy of this form would also affect the lensing shear of galaxies, in particular by generating a $B$-mode on large scales. This was investigated in \cite{adam_probing_2025}, which built on earlier work \cite{pitrou_weak-lensing_2015,pereira_weak-lensing_2016}, and produced forecasts for detectability of the effect by a Euclid-like survey.

Here we extend the analysis of \cite{adam_probing_2025} by considering the combination of galaxy shear with the lensing convergence of the CMB. We compute the signal-to-noise ratio (SNR) for the cross-correlation of the reconstructed CMB lensing convergence $\kappa_{\ell m}$ with the (tomographic) $B$-mode lensing shear $B_{\ell m}(z^i)$ that is generated by late-time anisotropic expansion, using the methods presented in \cite{adam_probing_2025}, and adopt the same notations and parameter values.

The paper is structured as follows. In \autoref{sec:Modelling_Anisotropy}, we describe the modelling of late-time, large-scale anisotropy using a Bianchi~I spacetime. \hyperref[sec:kappaxB_Cross_Correlation]{\Cref*{sec:kappaxB_Cross_Correlation}} reviews the observables relevant to the $\kappa$–$B$ cross-correlation, outlines the qualitative features of the resulting spectra in \autoref{sec:Spectrum_Features}, and presents signal-to-noise estimates in \autoref{sec:SNR_and_Sensitivity} for the cross-correlation between $B$-modes measured by the Euclid Wide Survey (EWS) and CMB lensing convergence from the Planck (Planck2018 and PlanckPR4) and Simons Observatory Large Aperture Telescope (SO LAT) surveys.


\section{Modelling large-scale anisotropy}\label{sec:Modelling_Anisotropy}

The Cosmological Principle consists of two independent assumptions that the Universe should adhere to on sufficiently large scales, namely, homogeneity and isotropy. Relaxing the latter assumption while retaining the first leads to the Bianchi models of spacetime, the simplest being Bianchi~I. This spacetime geometry exhibits different expansion rates along three orthogonal spatial directions. In the limit where these expansion rates are equal to one another, Bianchi~I models in particular reduce to spatially-flat FLRW spacetimes.

{Motivated by observational hints at a preferred direction,} we restrict ourselves to an axisymmetric Bianchi~I model, for which two of the expansion rates are equal. Nevertheless, most of the equations that we present do not rely upon this assumption. In coordinates where the spatial axes are aligned with the principal axes of expansion, the general form of the Bianchi~I line element is given by
\begin{equation}\label{eqn:Scale_Fact_First}
	\dd s^2 = -\dd t^2+a^2(t)\gamma_{ij}(t)\dd x^i\dd x^j = a^2(\eta)\left[-\dd \eta^2+\gamma_{ij}(\eta)\dd x^i\dd x^j\right],
\end{equation}
where cosmic time $t$ and conformal time $\eta$ are related in the usual manner: $a(\eta)\dd \eta=\dd t$.

In analogy to the conformal Hubble parameter $\H = a'/a$, which quantifies isotropic expansion, the (conformal) geometric shear $\bs{{\sigma}}$ is a measure of the rate of anisotropic expansion. This quantity is defined as\footnote{Note that this differs from the commonly-used definition of shear rate by a factor of $a$.}
\begin{equation}
	{\sigma}_{ij}= \frac{1}{2}{\gamma}_{ij}',
\end{equation}
where the prime represents a derivative with respect to conformal time $\eta$. The Einstein field equations for a Bianchi~I metric take the form \cite{pitrou_theory_2007}
\begin{subequations}\label{eqn:BI_EoM}
\begin{align}
		\H^2 &= \frac{8\pi G}{3} a^2\rho+\frac{1}{6} \sigma^2 \quad  
{\quad \left(\sigma^2 = \sigma_{ij} \sigma^{ij}\right) }\label{eqn:Friedmann_Bianchi_I} \\
		(\sigma\indices{^i_j})' &= -2\H\sigma\indices{^i_j}+ 8\pi G a^2\, \Pi\indices{^i_j},
  \label{eqn:Shear_EoM}	
\end{align}
\end{subequations}
where the total anisotropic stress $\Pi_{ij}$ present in the Universe drives the evolution of the shear $\bs{\sigma}$. For the anisotropic stress, we use the model \cite{pereira_weak-lensing_2016}
\begin{equation}
 	\Pi\indices{^i_j}(a) = f(a)W\indices{^i_j},
\end{equation}
where $W\indices{^i_j}$ is a constant {dimensionless} matrix. We use a simple form for $f$ which ties the anisotropic stress to dark energy: 
\begin{equation}
   { f(a) = {\rho_{c0}}\,{\Omega_{de}(a)},}
\end{equation}
where $\rho_{c0}=8\pi G/(3H_0^2)$ is the critical density.

\begin{figure}[htb]
	\centering
	\includegraphics[width=\WidthScale\linewidth]{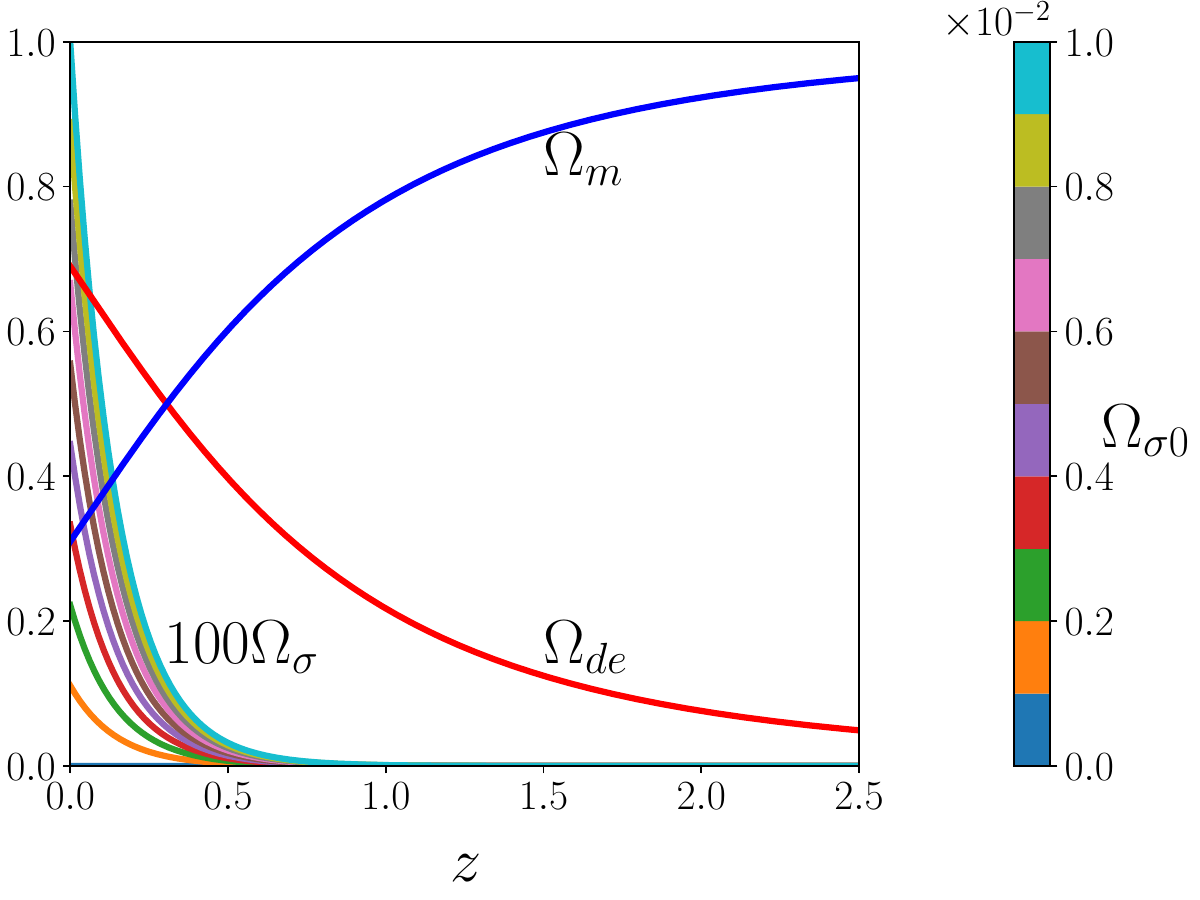 }
	\caption{Evolution of density parameters $\Omega_m$, $\Omega_{de}$ and $\Omega_{\sigma}$ (scaled by a factor of 100) up to $z=2.5$ for various values of $\Omega_{\sigma0}$.}
 \label{fig:Density_Params}
\end{figure}

In analogy with the usual matter density parameters, we define the shear density parameter
\begin{equation}
	 \Omega_\sigma = \frac{\sigma^2}{6\H^2},
\end{equation}
as a measure of the level of anisotropy in the Universe. The effect of varying the current shear density parameter $\Omega_{\sigma0}$ up to a maximum value of $10^{-2}$ is shown in \autoref{fig:Density_Params}. The procedure for generating the initial conditions and model parameters necessary to compute the evolution of $\bs{\sigma}$ is outlined in Appendix B of \cite{adam_probing_2025}, which also lists the values of the cosmological parameters used in this paper.

Since the observed Universe appears to be (mostly) well-described as isotropic, we expect the anisotropic expansion encoded by $\bs{\sigma}$ to be much weaker than the isotropic expansion described by the Hubble rate $\H$, so that 
\begin{equation}\label{eqn:sigma_H_ratio}
    \frac{|\sigma_{ij}|}{\H} \ll 1,
\end{equation}
for the entire history of the Universe, in order to be consistent with observations. Moreover, it is assumed that the spatial metric can be written as
\begin{equation}
	\gamma_{ij} \approx \delta_{ij}+2\beta_{ij},
\end{equation}
where $|\beta_{ij}|\ll 1$ are small, time-dependent, homogeneous perturbations to Euclidean 3-space, of the same order as $\sigma/\H$. In this regime, the shear is given by
\begin{equation}\label{eqn:sigma_weak_limit_beta}
    \sigma_{ij} = \beta_{ij}'.
\end{equation}

The ratio in \autoref{eqn:sigma_H_ratio} can therefore be treated as an additional perturbation parameter which couples to the standard FLRW scalar-vector-tensor perturbations. For our purposes, we expand observables to first order in $\sigma/\H$ in order to quantify the leading-order effects of late-time anisotropic expansion. For a more complete exposition on this perturbation scheme and its application to lensing observables, see \cite{pitrou_weak-lensing_2015,adam_probing_2025} and references therein. More general information on this approach to large-scale anisotropies and its application to the CMB can be found in \cite{pitrou_bianchi_2019, pontzen_linearization_2011, vicente_cmb_2025}.

Under the assumption of axisymmetry, the allowed form of the spatial shear $\bs{\sigma}$ (and anything derived from it) simplifies significantly. 
{If $\bm{e}$ is the unit direction of axisymmetry, then the shear is 
\begin{equation}
    \sigma_{ij}= \sigma_{\perp}\big(\delta_{ij} - 3e_ie_j\big) \equiv \sigma_{\perp} \hat{\Sigma}_{ij},
\end{equation}
where $\sigma_\perp$ is the shear rate transverse to $\bm{e}$ and the second equality defines the matrix $\hat{\bs{\Sigma}}$.}

This means that any quantity that is linear in the shear $\bs{\sigma}$ can be decomposed into a time-dependent function multiplied by the constant matrix $\hat{\bs{\Sigma}}$, which contains all necessary information about the principal directions of anisotropic expansion. For example, if we write $\beta_{ij} = \beta_\perp \hat{\Sigma}_{ij}$ and $W_{ij} = W_\perp \hat{\Sigma}_{ij}$ then \autoref{eqn:Shear_EoM} and \autoref{eqn:sigma_weak_limit_beta} simplify to
\begin{subequations}
    \begin{align}
        \beta_\perp'  &= \sigma_\perp\,,\\
        \sigma_\perp' &= -2\H \sigma_\perp + 8\pi G a^2 f W_\perp\,,
    \end{align}
\end{subequations}
while the density parameter becomes 
\begin{equation}
    \Omega_\sigma = \frac{\sigma_\perp^2}{\H^2}\, .
\end{equation}

Although we assume axisymmetry in our numerical calculations, we do not expect this to affect our conclusions. The majority of the equations we present are valid for any (weakly anisotropic) Bianchi~I spacetime. In particular, the estimator, covariance, and signal-to-noise expressions we present do not assume any special symmetry in the anisotropic expansion. Generally speaking, the only expressions which do assume axisymmetry will have a subscript $\perp$ somewhere within them. Under this assumption, $\sigma_\perp$ can be thought of as a proxy for the total anisotropic expansion strength. Since we are working to linear order in $\flatfrac{\sigma_{ij}}{\H}$, the magnitude of all observables (including signal-to-noise) will scale according to this shear strength.
\section{\boldmath Cross-correlation of CMB lensing and galaxy \texorpdfstring{$B$}{B}-mode shear}\label{sec:kappaxB_Cross_Correlation}

For the most part, the theoretical formulas for the $\kappa$-$B$ cross-correlations can be inferred from the $E$-$B$ case \cite{adam_probing_2025} with minimal alterations. Roughly speaking, the main differences are that the $\ell$-dependent shape factors change according to
\begin{align}
 \frac{(\ell+2)!}{(\ell-2)!}~~ \longmapsto ~~ \ell(\ell+1)\left[\frac{(\ell+2)!}{(\ell-2)!}\right]^{1/2} \,,   
\end{align}
and the convergence source distribution is now effectively a delta function at last scattering,
\begin{align}
 n^\kappa(z) = \delta_D(z-z_{\text{LS}})\quad \text{with} \quad z_{\text{LS}} \approx 1089 \,.   
\end{align}
Naturally, this means that tomographic information is only available for the $B$-modes and not for the CMB convergence multipoles.

To leading order in $\sigma/\H$, the non-vanishing $\kappa$-$B$ bipolar spherical harmonic ({\biposh}) coefficients are given by
\begin{equation}\label{eqn:EB_BipoSH_Coeff_Dominant}
    ^{\kappa B^i}\!\mathcal{A}^{2M}_{\ell, \ell\pm 1} = {\rm i}\,\frac{{^2F_{\ell \pm 1, 2, \ell}}}{4\sqrt{5}}\,\ell(\ell+1)\,\left[\frac{(\ell+2)!}{(\ell-2)!}\right]^{1/2}\,\mathcal{P}_{\ell M}^{\kappa i}, 
\end{equation}
where we have introduced the quantity
\begin{equation}\label{eqn:P_lM_defn}
    \mathcal{P}_{\ell M}^{\kappa i} = \frac{4}{\ell(\ell+1)}\,\left[\frac{(\ell-2)!}{(\ell+2)!}\right]^{1/2} \int \dd k\, k^2 P(k) \Delta^\kappa_\ell (k) \Delta^i_{\ell M} (k)\, . 
\end{equation}
Here $P$ is the power spectrum of the primordial curvature perturbation $\mathcal{R}$, with the convention $\expval{\mathcal{R}(\bm{k})\mathcal{R}(\bm{k}')^{*} } = P(k)\delta(\bm{k}-\bm{k}')$. 
The $\ell$-dependent prefactors are \cite{hu_weak_2000,pitrou_weak-lensing_2015, adam_probing_2025}
\begin{subequations}
	\begin{align}
        ^2F_{\ell+1, 2, \ell} &= (-1)^{\ell+1}(\ell-2)\bigg[{\frac{15}{\pi}}\,{\frac{(\ell-1)(\ell+3)}{\ell(\ell+1)(\ell+2)}}\bigg]^{1/2} \\
        ^2F_{\ell-1, 2, \ell} &= (-1)^{\ell+1}(\ell+3)\bigg[{\frac{15}{\pi}}\,{\frac{(\ell-2)(\ell+2)}{\ell(\ell-1)(\ell+1)}}\bigg]^{1/2}\, .
	\end{align}
\end{subequations}
For large $\ell$, \autoref{eqn:EB_BipoSH_Coeff_Dominant} has the scaling behaviour
\begin{align}
  ^{\kappa B^i}\!\mathcal{A}^{2M}_{\ell, \ell\pm 1} \sim \ell^{4.5}\, \mathcal{P}_{\ell M}^{\kappa i}\,.  
\end{align}

The $\ell$-dependent kernels are defined in terms of the spherical Bessel function $j_\ell$ and the Weyl potential transfer function $T_\varphi$:
\begin{subequations}
    \begin{align}        
        \Delta_\ell^\kappa(k) &= \frac{1}{2}\ell(\ell+1)\sqrt{\frac{2}{\pi}}\int_0^{\chi_*} \dd\chi \,q^\kappa(\chi,\chi_*)\,j_\ell(k\chi)\,T_\varphi(\chi,k)\,, \\
        \Delta^{i}_{\ell M}(k) &= \frac{1}{2}\bigg[{\frac{2}{\pi}}\,{\frac{(\ell+2)!}{(\ell-2)!}}\bigg]^{1/2}\int_{0}^{\chi_S}\dd\chi \, q^i(\chi,\chi_S)\,j_{\ell}(k\chi) \,\alpha_{2 M}(\chi)\,T_\varphi(\chi,k).
    \end{align}
\end{subequations}
We define the Weyl potential $\varphi$ and its transfer function through $\varphi(\eta, \bm{k}) = \Phi(\eta, \bm{k}) + \Psi(\eta, \bm{k}) = T_\varphi(\eta,k)\mathcal{R}(\bm{k})$, where $\Phi$ and $\Psi$ are the usual Bardeen potentials and $\mathcal{R}$ is the primordial curvature perturbation.
The quadrupole coefficients $\alpha_{2M}$ quantify the angular deflection induced by the Bianchi~I degrees of freedom. More precisely, they represent the multipole moments of an effective Bianchi~I `lensing potential,' whose derivatives determine the deflection angle arising from anisotropic expansion \cite{pitrou_weak-lensing_2015}. In the case of axisymmetric anisotropy, the multipoles can be written $\alpha_{2M}=\alpha_\perp \hat{\Sigma}_{2M}$,
where we have introduced the deflection strength
\begin{equation}
    \alpha_\perp(\chi) = -\frac{2}{\chi}\int_0^\chi\dd\tilde{\chi}\,\beta_\perp(\tilde{\chi}),
\end{equation}
and the quadrupolar coefficients $\hat{\Sigma}_{2M}$ are defined in terms of the matrix $\hat{\bs{\Sigma}}$ through
\begin{equation}\label{eqn:Sigma_2m_Multipoles}
    \hat{\Sigma}_{2M} = 
    \left(\frac{\pi}{30}\right)^{1/2}
    \begin{cases}	\sqrt{6}\,\big(\hat{\Sigma}_{xx}+\hat{\Sigma}_{yy}\big) & \quad M=0\,, \\
    2 \,\big(\mp\hat{\Sigma}_{xz}+{\rm i}\,\hat{\Sigma}_{yz}\big)  &  \quad M=\pm 1 \,,\\
    \hat{\Sigma}_{xx}-\hat{\Sigma}_{yy} \mp 2{\rm i}\,\hat{\Sigma}_{xy} &  \quad M=\pm 2\,. 
 \end{cases}
\end{equation}
The $M$ index  contains the five degrees of freedom necessary to reconstruct the directions and rates of anisotropic expansion. Note, however, that in our numerical simulations we impose axisymmetry about the $z$-axis so that only the $M=0$ multipole is non-zero.

The effective lensing efficiencies are simply
\begin{subequations}
\begin{align}
    q^\kappa(\chi) &= \frac{\chi^*-\chi}{\chi^* \chi} \\
    {q}^i(\chi) &= \int_\chi^{\chi_s}\text{d}\tilde{\chi}\,\frac{\tilde{\chi}-\chi}{\tilde{\chi}\chi}\mathcal{N}^i(\tilde{\chi}),
\end{align}
\end{subequations}
where $\mathcal{N}_i(\chi) = n_i(z(\chi)){\text{d}z}/{\text{d}\chi} = n_i(z(\chi))H(\chi)$ is the source distribution as a function of conformal distance $\chi$, $\chi^*$ is the conformal distance to last scattering, and $\chi_s$ is the maximum distance probed by the lensing survey. If we make use of the Limber approximation and recognise that $q^i(\chi)\approx 0$ for $\chi > \chi_s$, \autoref{eqn:P_lM_defn} reduces to
\begin{equation}\label{eqn:P_lM_defn_Limber}
	\mathcal{P}_{\ell M}^{\kappa i} \approx \int_{0}^{\chi_s} \frac{\dd\chi}{\chi^{2}}\, q^{\kappa}(\chi)\,q^{i}(\chi) P\left(\frac{\nu_\ell}{\chi}\right)\alpha_{2M}(\chi)\left|T_{\varphi}\left(\chi,\frac{\nu_\ell}{\chi}\right)\right|^{2}\,,
\end{equation}
with $\nu_\ell = \ell + 1/2$.

The CMB lensing convergence power spectrum is given by 
\begin{equation}
    C_\ell^{\kappa\kappa} = \int \dd k \, k^2 P(k) \Delta_\ell^\kappa(k)^2.
\end{equation}
Applying the Limber approximation leads to the standard result
\begin{equation}\label{eqn:Cl_kappa_Limber}
    C_\ell^{\kappa\kappa} \approx  \frac{1}{4} \ell^2 (\ell+1)^2\int_{0}^{\chi_*}\frac{\dd \chi}{\chi^2} \, \,P\left(\frac{\nu_\ell}{\chi}\right) \left|q^{\kappa}(\chi,\chi_*)\,T_\varphi\left(\chi,\frac{\nu_\ell}{\chi}\right)\right|^{2}.
\end{equation}

We make use of the Limber approximation in order to compute all theoretical angular power spectra and {BipoSH} coefficients. 
This approximation is well-suited to the broad, slowly-varying kernels found in lensing and should be fairly accurate for lensing observables. In the standard lensing case, the error introduced by the Limber approximation is smaller than cosmic variance for scales larger than $\ell \gtrsim 10$ \cite{kilbinger_precision_2017}. Since the integrands involved in our case are similar to those in standard lensing, we expect similar behaviour. Moreover, as we show in \autoref{sec:SNR_and_Sensitivity}, cosmic variance leads to large uncertainties and low signal-to-noise on large scales. 
Therefore, we expect minimal information loss on the largest scales where Limber is least accurate. Nevertheless, going beyond the Limber approximation is relatively straightforward should one wish to calculate the low-$\ell$ modes more accurately.



\subsection{\boldmath Features of the {cross-power} spectra}\label{sec:Spectrum_Features}
    
In order to understand the general behaviour and features of $\mathcal{P}_{\ell M}^{\kappa i}$, we need to analyse the quantities in \autoref{eqn:P_lM_defn_Limber}. This will give us a better qualitative and quantitative grasp of this observable and its constraining power. Note that we plot $ ^{\kappa B^i}\!\mathcal{A}^{2M}_{\ell, \ell\pm 1} \sim \ell^{4.5} \, \mathcal{P}_{\ell M}^{\kappa i}$ rather than $\mathcal{P}_{\ell M}^{\kappa i}$, as the {\biposh} is mostly flat for low $\ell$ and better illustrates the effects of redshift and anisotropy strength.

\begin{figure}[htb!]
    \centering
    \includegraphics[width=\WidthScale\linewidth]{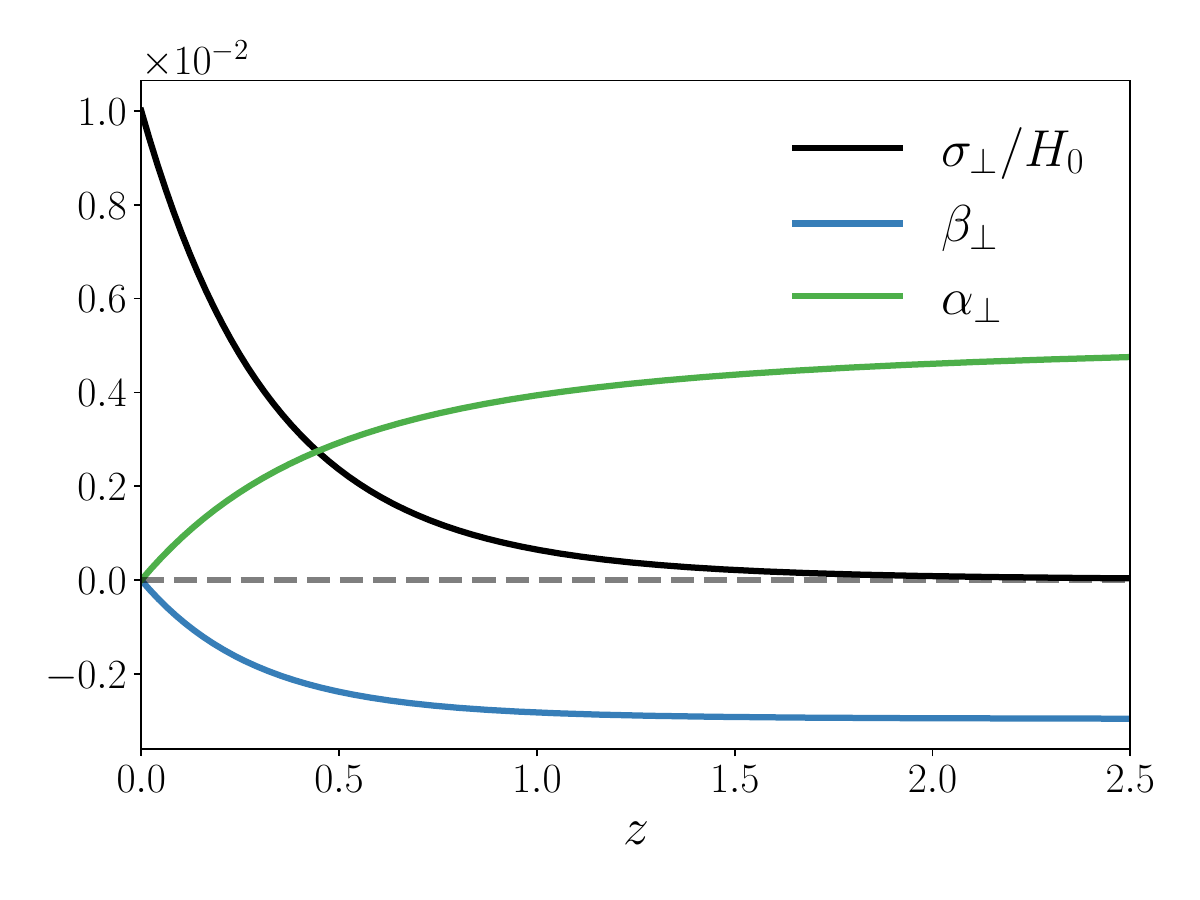}
    \caption{Evolution of $\alpha_\perp$, $\beta_\perp$, and $\sigma_{\perp}/H_0$ up to $z=2.5$ for a final shear strength of $\Omega_{\sigma0} = 10^{-4}$ (i.e. $\sigma_{\perp0}/H_0 = 10^{-2}$).
    }
    \label{fig:alpha_beta_sigma_z}
\end{figure}

The principal difference between the integrands in \autoref{eqn:P_lM_defn_Limber} and \autoref{eqn:Cl_kappa_Limber} is the appearance of the $\alpha_{2M}$ quadrupoles in the former. The behaviour of the deflection strength $\alpha_\perp$ as a function of redshift, along with the corresponding shear $\sigma_\perp$ and metric perturbation $\beta_\perp$, is shown in \autoref{fig:alpha_beta_sigma_z}. Since the shear is only present at late times, both $\beta_\perp$ and $\alpha_\perp$ flatten out at larger $z$. Intuitively, light rays emitted by more distant sources only encounter large-scale anisotropy at lower redshifts, leading to a roughly uniform amount of lensing from these late-time effects. The integral in \autoref{eqn:P_lM_defn_Limber} thus receives more contributions from higher redshifts. Since $k \sim \ell/\chi(z)$, small-scale and late-time features of the matter power spectrum are effectively downweighted. As we shall see in \autoref{sec:SNR_and_Sensitivity}, this shifts much of the signal and constraining power to lower $\ell$. Different models of late-time anisotropic expansion will affect any resulting angular spectra by changing the rate at which $\alpha_\perp$ levels off as a function of $z$. This makes distinguishing between models challenging, since $\alpha_\perp$ --- being a twice-integrated version of the shear --- smooths out the features of $\sigma_\perp$. Nevertheless, the tomographic spectra we investigate are sensitive to both the amplitude of anisotropy (i.e. $\Omega_{\sigma 0}$ or $\sigma_{\perp0}/H_0$) and the approximate redshift at which it becomes relevant.

\begin{figure}[htb!]
    \centering
    \includegraphics[width=\linewidth]{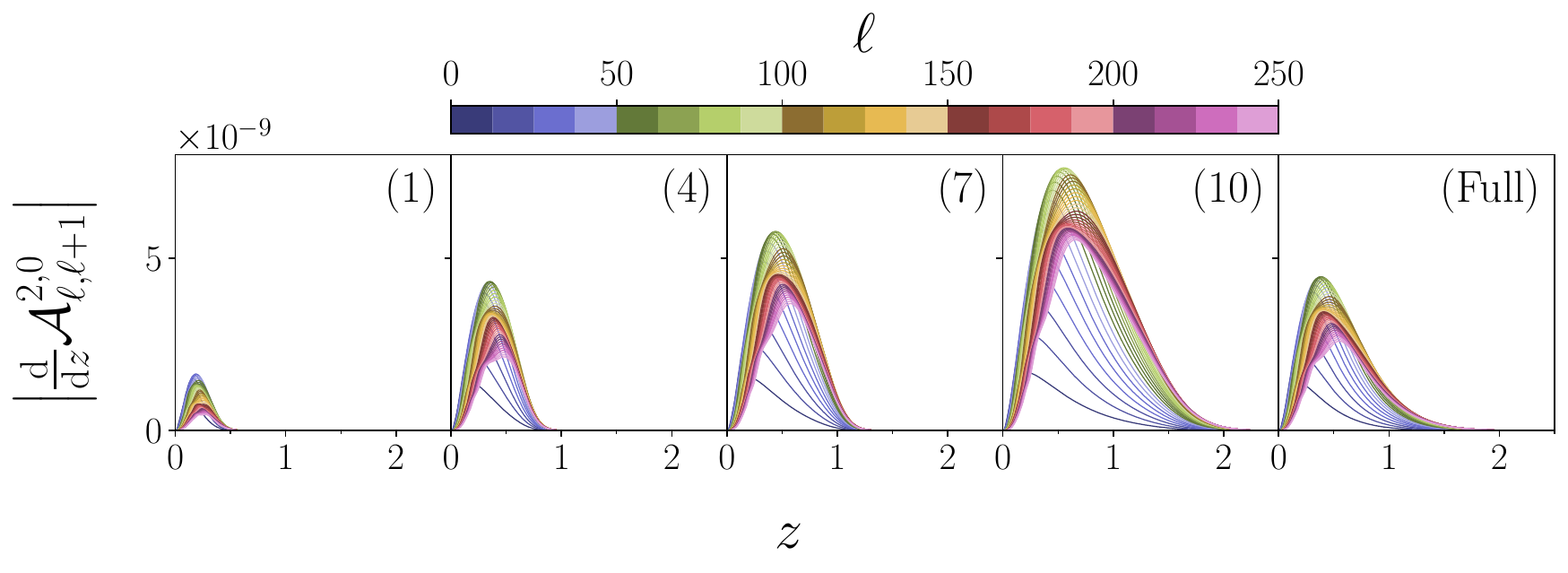}
    \caption{
        Integrand of {$\mathcal{P}_{\ell M}^{\kappa i}$, given by} \autoref{eqn:P_lM_defn_Limber}, scaled by the $\ell$-dependent prefactors from \autoref{eqn:EB_BipoSH_Coeff_Dominant}, as a function of redshift up to $z = 2.5$ for a final shear strength of $\Omega_{\sigma 0} = 10^{-4}$ (i.e., $\sigma_{\perp0} / H_0 = 10^{-2}$). Curves are shown for four representative redshift bins, as well as for the full (non-tomographic) redshift distribution, over the range $10 \leq \ell \leq 250$. Note that changing variables in \autoref{eqn:P_lM_defn_Limber} from $\chi$ to $z$ introduces a factor of $H^{-1}=(1+z)^{-1}\H^{-1}$ into each integrand.
    }
    \label{fig:A2M_lp1_Integrand}
\end{figure}
\autoref{fig:A2M_lp1_Integrand} shows the integrand of the {\biposh} coefficient $^{\kappa B^i}\!\mathcal{A}^{2,0}_{\ell,\ell+1}$ for a selection of redshift bins and a range of $\ell$ values. The plotted quantity, $\dd \big({^{\kappa B^i}\!\mathcal{A}^{2,0}_{\ell,\ell+1}}\big)/{\dd z}$, is defined such that the area under each coloured curve (over $0 \leq z \leq 2.5$) equals the corresponding {\biposh} coefficient $^{\kappa B^i}\!\mathcal{A}^{2,0}_{\ell,\ell+1}$. The analogous plot for the $^{\kappa B^i}\!\mathcal{A}^{2,0}_{\ell, \ell - 1}$ coefficient is not shown since it is visually indistinguishable from the $\ell+1$ case. Evidently, the integrands grow in amplitude with $\ell$ until reaching a maximum around $30 \lesssim \ell \lesssim 100$ (as seen in the blue and green curves), indicating a turnover scale for $^{\kappa B^i}\!\mathcal{A}^{2,0}_{\ell, \ell + 1}$ in this range. Not unexpectedly, the curves also increase in both width and height as a function of redshift. All curves, regardless of redshift or $\ell$ value, become negligibly small well before $z=2.5$, indicating that the truncation of the upper limit of the integral in \autoref{eqn:P_lM_defn_Limber} is a good approximation.

\begin{figure}[htb!]
    \centering
    \includegraphics[width=\linewidth]{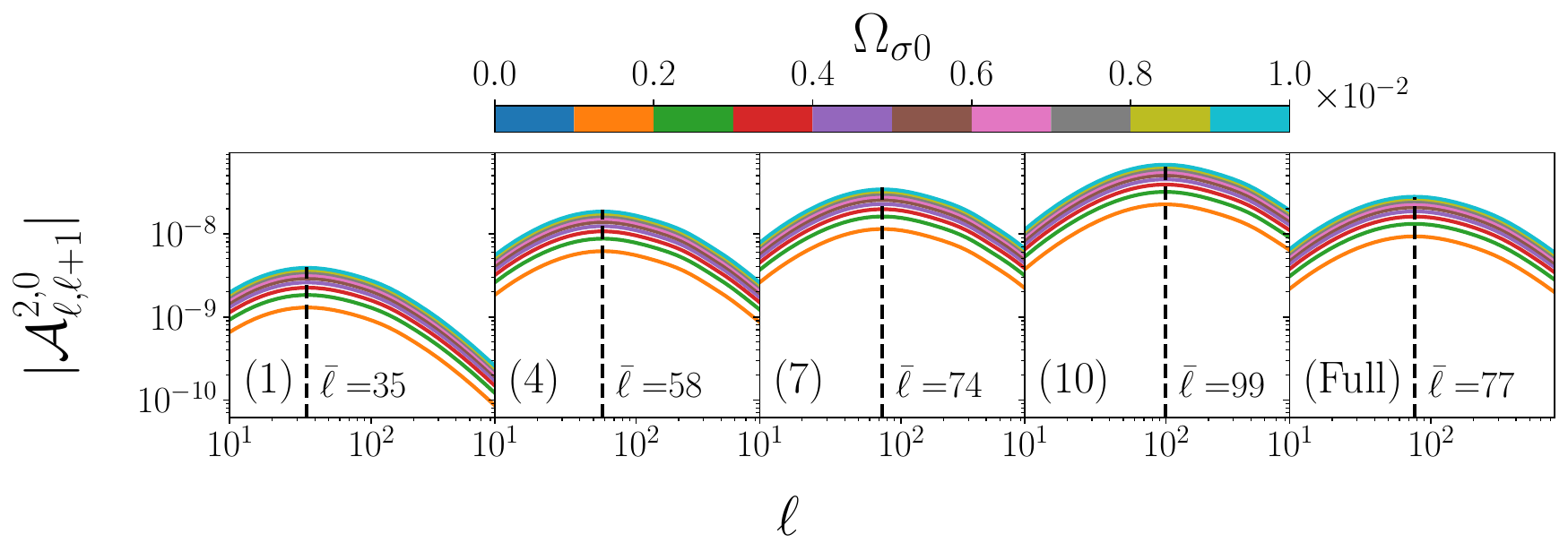}
    \caption{
        Amplitude of the {\biposh} coefficient $^{\kappa B^i}\!\mathcal{A}^{2,0}_{\ell, \ell + 1}$ over the range $10 \leq \ell \leq 750$, shown for the same redshift bins and distributions as in \autoref{fig:A2M_lp1_Integrand}. Different colours indicate 10 values of the shear strength in the range $\Omega_{\sigma 0} \leq 10^{-2}$ (i.e., $\sigma_{\perp0} / {H}_0 \leq 10^{-1}$). The turnover scale $\bar{\ell}$ for each redshift bin is indicated with a vertical dashed line. 
    }
    \label{fig:A2M_lp1}
\end{figure}

From \autoref{fig:A2M_lp1}, we see that the amplitudes of the {\biposh} coefficients increase with both redshift and anisotropy strength. Within each bin, the turnover scale $\bar{\ell}$ is unaffected by the anisotropy parameter $\Omega_{\sigma 0}$. At least for the model presented here, $\sigma_{\perp0}/H_0$ acts as an overall scaling factor for the curves. Consequently, the overall shapes of the spectra are insensitive to the magnitude of late-time anisotropy. In principle, by analysing the redshift evolution of the magnitude, one could determine when the shear $\bs{\sigma}$ begins to grow and become relevant. Although one may have to consider finer redshift bins at late time in order to achieve this.

    \subsection{Estimator and covariance}

Suppose that we have a set of convergence multipoles $\hat{\kappa}_{\ell m}$ which have been reconstructed from an observed map of CMB temperature fluctuations. Furthermore, let $\hat{B}^i_{\ell m}$ denote the $B$-mode multipoles for the $i$th redshift bin obtained from a weak-lensing survey's catalogue of galaxy shapes. Provided that the reconstructed convergence noise and $B$-mode shear noise are uncorrelated, an unbiased estimator of the $\kappa$-$B$ {\biposh} coefficients for the $i$th redshift bin is 
\begin{equation}\label{eqn:kappaB_BipoSH_Estimator}
    ^{\kappa B^i}\!\hat{\A}^{LM}_{\ell \ell'} = \sqrt{2L+1} \largesum_{m,m'}(-1)^{L+m} \mqty(\ell & \ell' & L \\ -m & m' & M) \hat{\kappa}_{\ell m}\hat{B}^{i*}_{\ell' m'}\,.
\end{equation}
By inverting \autoref{eqn:EB_BipoSH_Coeff_Dominant}, we can construct the {simple} estimator 
\begin{equation}\label{eqn:P_lM_Estimator}
    \hat{\mathcal{P}}_{\ell M}^{\kappa i} = -2\sqrt{5}\,{\rm i}\,\frac{1}{\ell(\ell+1)}\,\left[{\frac{(\ell-2)!}{(\ell+2)!}}\right]^{1/2}\largesum_{I=\pm 1}\frac{^{\kappa B^i}\!\hat{\A}^{2M}_{\ell, \ell+I}}{{^2F_{\ell+I, 2, \ell}}}\,,
\end{equation}
which weights the $\ell+1$ and $\ell-1$ contributions equally.

Assuming statistical isotropy, 
i.e. neglecting correlations of the CMB convergence $\kappa$ and the shear $B$-mode, and assuming Gaussian fluctuations,
the covariance of the estimator $\hat{\mathcal{P}}_{\ell M}^{\kappa i}$ can be approximated as
\begin{equation}\label{eqn:P_lM_Estimator_Covariance}
\text{Cov}\left(\hat{\mathcal{P}}_{\ell M}^{\kappa i}, \hat{\mathcal{P}}_{\ell' M'}^{\kappa j}\right)_{\text{SI}} = \frac{20}{f_{\text{sky}}} \left[ \frac{\ell^2 (\ell+1)^2(\ell+2)!}{(\ell-2)!} \right]^{-1} \largesum_{I=\pm 1}\frac{{\left(C^{\kappa \kappa}_{\ell}\right)_{\text{SI}}} {\left(C^{B^i B^j}_{\ell+I}\right)_{\text{SI}}}}{\left({^2F_{\ell+I, 2 ,\ell}}\right)^2} \delta_{\ell \ell'}\delta_{MM'},
\end{equation}
where $f_{\text{sky}}$ is the fraction of sky covered by the overlap between the CMB and weak-lensing surveys' footprints, and the statistically-isotropic angular power spectra are defined through
\begin{subequations}
    \begin{align}
        \expval{\hat{\kappa}_{\ell m} \hat{\kappa}^{*}_{\ell' m'}}_{\text{SI}} &= {\left(C^{\kappa\kappa}_{\ell}\right)_{\text{SI}}}\delta_{\ell \ell'}\delta_{m m'} = \left(C^{\kappa\kappa}_\ell + N^{\kappa\kappa}_\ell\right)\delta_{\ell \ell'}\delta_{m m'}\,, \\
        \expval{\hat{B}^i_{\ell m} \hat{B}^{j*}_{\ell' m'}}_{\text{SI}} &= {\left(C^{B^i B^j}_{\ell}\right)_{\text{SI}}}\delta_{\ell \ell'}\delta_{m m'} = \frac{\expval{\gamma_{\text{int.}}^2}}{\bar{N}_{i}}\delta_{ij}\delta_{\ell \ell'}\delta_{m m'}\,, \\        
        \expval{\hat{\kappa}_{\ell m} \hat{B}^{i*}_{\ell' m'}}_{\text{SI}} &= {\left(C^{\kappa B^i}_{\ell}\right)_{\text{SI}}}\delta_{\ell \ell'}\delta_{m m'} = 0\,.
    \end{align}
\end{subequations}
The $B$-mode variance is estimated as pure shape noise with an intrinsic ellipticity variance of $\expval{\gamma_{\text{int.}}^2}=0.3^2$ (as in \cite{blanchard_euclid_2020,deshpande_euclid_2024}), while the reconstructed CMB convergence noise $N^{\kappa\kappa}_\ell$ is survey specific. Euclid is expected to observe a galaxy source density of around $\bar{N}=30$ arcmin$^{-2}$ which, when distributed among 10 equi-populated bins, yields $\bar{N}_i=3$ arcmin$^{-2}$ \cite{blanchard_euclid_2020,deshpande_euclid_2024}.

\begin{figure}[htb!]
    \centering
    \includegraphics[width=\WidthScale\linewidth]{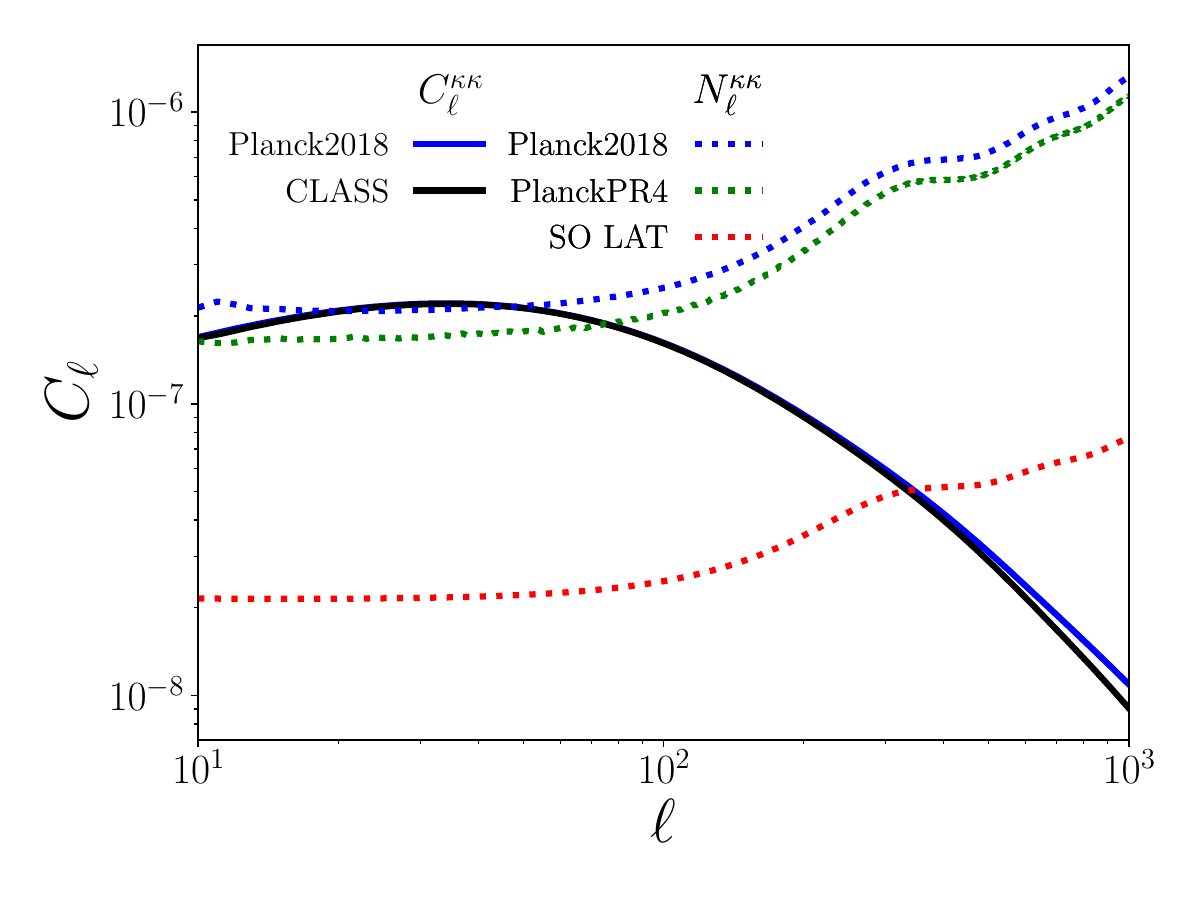}
    \caption{Linear convergence power spectrum computed using CLASS (solid black) alongside the Planck2018 minimum-variance power spectrum (solid blue). The minimum-variance reconstruction noise/bias spectra for Planck2018 (dotted blue) and Simons (dotted red) are also shown.}
    \label{fig:Convergence_Spectra_and_Noise}
\end{figure}

As can be seen from \autoref{fig:Convergence_Spectra_and_Noise}, the linear CMB convergence spectrum calculated using CLASS \cite{lesgourgues_cosmic_2011} closely matches the convergence spectrum obtained from Planck 2018 data using a minimum-variance estimator on all scales of interest. The lensing reconstruction noise biases for both Planck\footnote{\label{fn:PLANCK}The Planck 2018 lensing power spectrum and bias can be found in the \texttt{COM\_Lensing\_4096\_R3.00} lensing data package obtained from the \href{https://pla.esac.esa.int}{Planck Legacy Archive}. For Planck PR4, we use an empirical estimate of the bias kindly provided to us by Julien Carron.} and Simons Observatory\footnote{\label{fn:SO}For Simons Observatory, we make use of the minimum-variance lensing bias obtainable from \href{https://github.com/simonsobs/so_noise_models/blob/master/LAT_lensing_noise/lensing_v3_1_0/nlkk_v3_1_0deproj0_SENS2_fsky0p4_it_lT30-3000_lP30-5000.dat}{this link}.} have a characteristic increase at large $\ell$. This results in strongly suppressed SNR values for our estimator $\hat{\mathcal{P}}_{\ell M}^{\kappa i}$ on small angular scales.

\begin{table}[htb!]
    \centering
    \begin{subtable}[t]{0.55\textwidth}
        \centering
        \begin{tabular}{@{}rlll@{}}
\toprule
                             & \multicolumn{1}{c}{\textbf{Planck2018}}     & \multicolumn{1}{c}{\textbf{PlanckPR4}}             & \multicolumn{1}{c}{\textbf{SO LAT}}   \\ \midrule
$\bs{f_{\textbf{sky}}}$      & $0.32$                                       & $0.32$                                              & $0.22$                                \\
$\bs{N_\ell^{\kappa\kappa}}$ & MV estimate\textsuperscript{\ref{fn:PLANCK}} & Empirical estimate\textsuperscript{\ref{fn:PLANCK}} & MV model\textsuperscript{\ref{fn:SO}} \\ \bottomrule
\end{tabular}
        \caption{CMB lensing survey inputs.}
        \label{tbl:CMB_Surveys}
    \end{subtable}
    \hfill
    \begin{subtable}[t]{0.35\textwidth}
        \centering
        \begin{tabular}{@{}rl@{}}
        \toprule
        \multicolumn{2}{c}{\textbf{EWS}}                        \\ \midrule
        $\bs{\langle \gamma_{\textbf{int.}}^2 \rangle}$ & $0.3^2$ \\
        $\bs{\bar{N}}$ [arcmin$^{-2}$]                & 30      \\ \bottomrule
        \end{tabular}
        \caption{Galaxy weak-lensing inputs.}
        \label{tbl:WL_Survey}
    \end{subtable}
    \caption{
    Summary of input parameters for (a) CMB lensing (Planck, Simons) and (b) galaxy weak-lensing (Euclid) surveys. The bias $N_\ell^{\kappa\kappa}$ for each CMB survey corresponds to that of a convergence estimator constructed from a minimum-variance (MV) combination of temperature and polarisation multipoles. When making use of tomography, the sources are split evenly between 10 bins for a source density of $\bar{N}_i=3$ arcmin$^{-2}$.}
    \label{tbl:Surveys}
\end{table}

For $f_{\text{sky}}$ we use an estimate of the fraction of sky covered by the overlap between the footprint of the Euclid Wide Survey (EWS) and the footprint of the CMB survey in question. Since the Planck satellite was able to survey most of the sky, we use the Euclid sky coverage of $f_{\text{sky}} = 0.32$ \cite{collaboration_euclid_2025-1}. For the cross-correlation between the Simons  and Euclid surveys, we estimate the overlapping footprint area to be approximately 9000 deg$^2$ (corresponding to $f_{\text{sky}} \approx 0.22$), based on Fig. 3 of \cite{collaboration_simons_2025}. 

A summary of all survey inputs used in this investigation can be found in \autoref{tbl:Surveys}.

    \subsection{Signal-to-noise and sensitivity}\label{sec:SNR_and_Sensitivity}

Since the covariance in \autoref{eqn:P_lM_Estimator_Covariance} is completely diagonal (i.e., $\propto \delta_{\ell \ell'}\delta_{M M'}\delta_{ij}$), we can construct measures of the approximate signal-to-noise ratio (SNR) for our estimator using expressions of the form
\begin{subequations}
    \begin{align}
        \left|\frac{{\mathcal{P}}_{\ell M}^{\kappa}}{\Delta {\mathcal{P}}_{\ell M}^{\kappa}}\right|^2 &= \largesum_{i} \left(\frac{{\mathcal{P}}_{\ell M}^{\kappa i}}{\Delta {\mathcal{P}}_{\ell M}^{\kappa i}}\right)^2 ,\label{eqn:SNR_Sum_Tomography}\\
        \left(\frac{S}{N}\right)^2_{\ell_{\text{max}}} &=  \largesum_{\ell = \ell_{\text{min}}}^{\ell_{\text{max}}}\sum_{M} \left|\frac{{\mathcal{P}}_{\ell M}^{\kappa}}{\Delta {\mathcal{P}}_{\ell M}^{\kappa}}\right|^2,
    \end{align}
\end{subequations}
where $\Delta \mathcal{P}^{\kappa i}_{\ell M}= \big[{\text{Var}\big(\hat{\mathcal{P}}_{\ell M}^{\kappa i}\big)_{\text{SI}}}\big]^{1/2}$. In the case where the tomographic information from a lensing survey is not used, the sum over the redshift bins $i$ drops out of \autoref{eqn:SNR_Sum_Tomography}.

For the purposes of modelling tomography, we make use of the ten equi-populated bins used by Euclid in their forecasting \cite{blanchard_euclid_2020, deshpande_euclid_2024}. The tomographic source distributions for these bins $n_i(z)$ are constructed by weighting the underlying source distribution $n(z)$ by the probability that a galaxy detected in that bin actually has a measured photometric redshift within the bin's specified redshift range. This process is also outlined in Appendix E of \cite{adam_probing_2025} and a plot of the resulting distributions is shown in \autoref{fig:Euclid_Tomographic_Distributions}.

As in \cite{adam_probing_2025}, we choose our limiting scales $\ell_{\text{min}}$ and $\ell_{\text{max}}$ in order to avoid problematic systematics and nonlinearities which may occur outside of this scale range. We set $\ell_{\text{min}} = 10$ since systematics and the breakdown of the Limber approximation can lead to issues on these largest scales. By considering \autoref{fig:SNR_Tomo_Full} where HaloFit nonlinear corrections have been included, we see that choosing $\ell_{\text{max}} = 200$ eliminates the worst regions of nonlinear contamination.

\begin{figure}[htb!]
    \centering
    \includegraphics[width=0.9\linewidth]{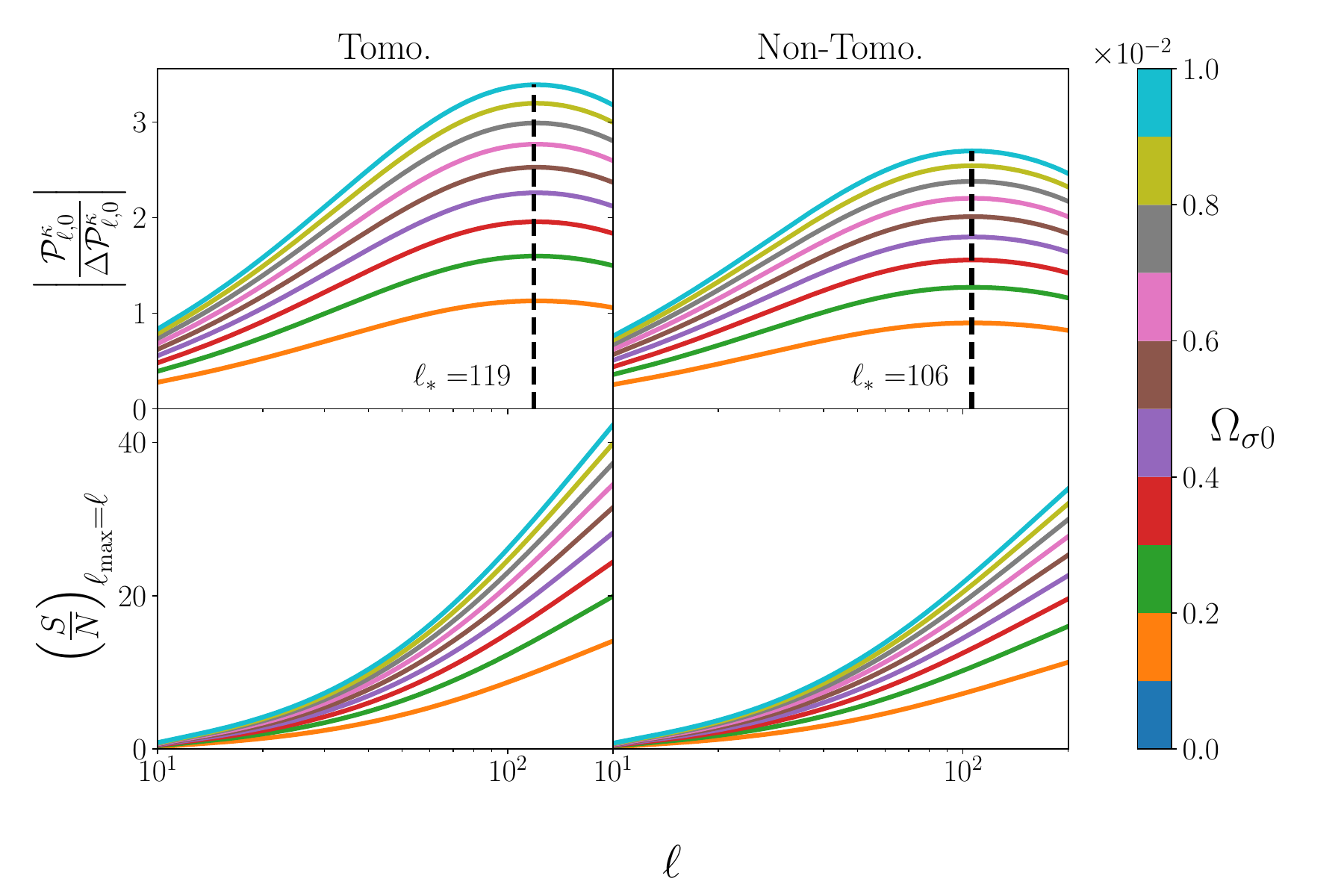}
    \caption{Individual (top row) and cumulative (bottom row) SNRs for $\hat{\mathcal{P}}^{\kappa}_{\ell,0}$ computed for Simons\,$\times$\,Euclid as a function of multipole moment $\ell$. The left column shows results summed over all tomographic bins, while the right column shows the corresponding non-tomographic results. Each curve corresponds to one of ten values of $\Omega_{\sigma 0} \leq 10^{-2}$. The scale of maximum individual SNR ($\ell_*$) is indicated with a vertical dashed line for both the tomographic and non-tomographic cases.}
    \label{fig:SNR_Tomo_Full_Individual_Sum}
\end{figure}

For the Simons\,$\times$\,Euclid cross-correlation, the difference between the tomographic and non-tomographic SNRs is illustrated in \autoref{fig:SNR_Tomo_Full_Individual_Sum}. This should be contrasted with \autoref{fig:SNR_Tomo_Full}, which shows the SNR for each redshift bin individually. Although the SNR for $\hat{\mathcal{P}}^{\kappa}_{\ell M}$ may be insufficient to constrain $\Omega_{\sigma 0}$ within individual redshift bins or at specific multipoles, combining information from all bins and multipoles will certainly increase the overall SNR and thereby improve sensitivity to the anisotropy parameter $\Omega_{\sigma 0}$. 

The SNRs for each individual multipole shown in the top row of \autoref{fig:SNR_Tomo_Full_Individual_Sum} exhibit a characteristic peak around $\ell\sim 100$. On larger scales, the signal is dominated by cosmic variance, whereas on smaller scales, the convergence reconstruction bias (see \autoref{fig:Convergence_Spectra_and_Noise}) significantly suppresses the SNR. The peak lies between these two regions. As noted in \autoref{sec:Spectrum_Features}, the shear anisotropy parameter $\Omega_{\sigma 0}$ acts as an overall scaling factor and so does not change the overall shape of the SNR or the scale at which it is maximised. 

The SNR for $\hat{\mathcal{P}}^{\kappa}_{\ell M}$ therefore peaks at a much larger scale than that of the usual $E$-mode power spectrum. As shown in \autoref{fig:C^EE_l_SNR}, the maximum SNR for $C^{EE}_\ell$ occurs near $\ell \sim 1100$, well within the nonlinear regime.

From \autoref{fig:SNR_Tomo_Full}, it is clear that the SNR peak shifts to smaller scales at higher redshifts. Heuristically speaking, this is because spatial Fourier modes $k$ are projected onto angular modes $\bar{\ell}$ satisfying $k\sim \bar{\ell}/\chi(\bar{z})$, where $\bar{z}$ is the mean redshift of a sample. $\P^{\kappa i}_{\ell M}$ is a projection of the power spectrum $P(k,z)=P(k)T^2_\varphi(k,z)$ and so will receive the bulk of its power from scales where $T_\varphi$ is constant --- i.e. $ k \lesssim k_{\text{eq}}$, the matter-radiation equality scale. Any features in $\P^{\kappa i}_{\ell M}$ and its SNR will therefore generally shift to smaller scales as redshift increases. Of course, this argument applies to any redshift-dependent angular spectrum.

Increasing the depth of a survey to a higher redshift would increase the SNR and improve constraining ability. Since we are modelling late-time anisotropy, any light beams originating from high-redshift sources experience the same level of distortion due to anisotropic modes. We therefore expect the SNR to saturate at some redshift and any information gain beyond this point should be minimal. As evidenced by \autoref{fig:SNR_Tomo_Full} and \autoref{fig:SNR_Var_Bins}, however, this saturation point lies beyond the maximum redshift of Euclid. 

\begin{figure}[htb!]
    \centering
    \includegraphics[width=\WidthScale\linewidth]{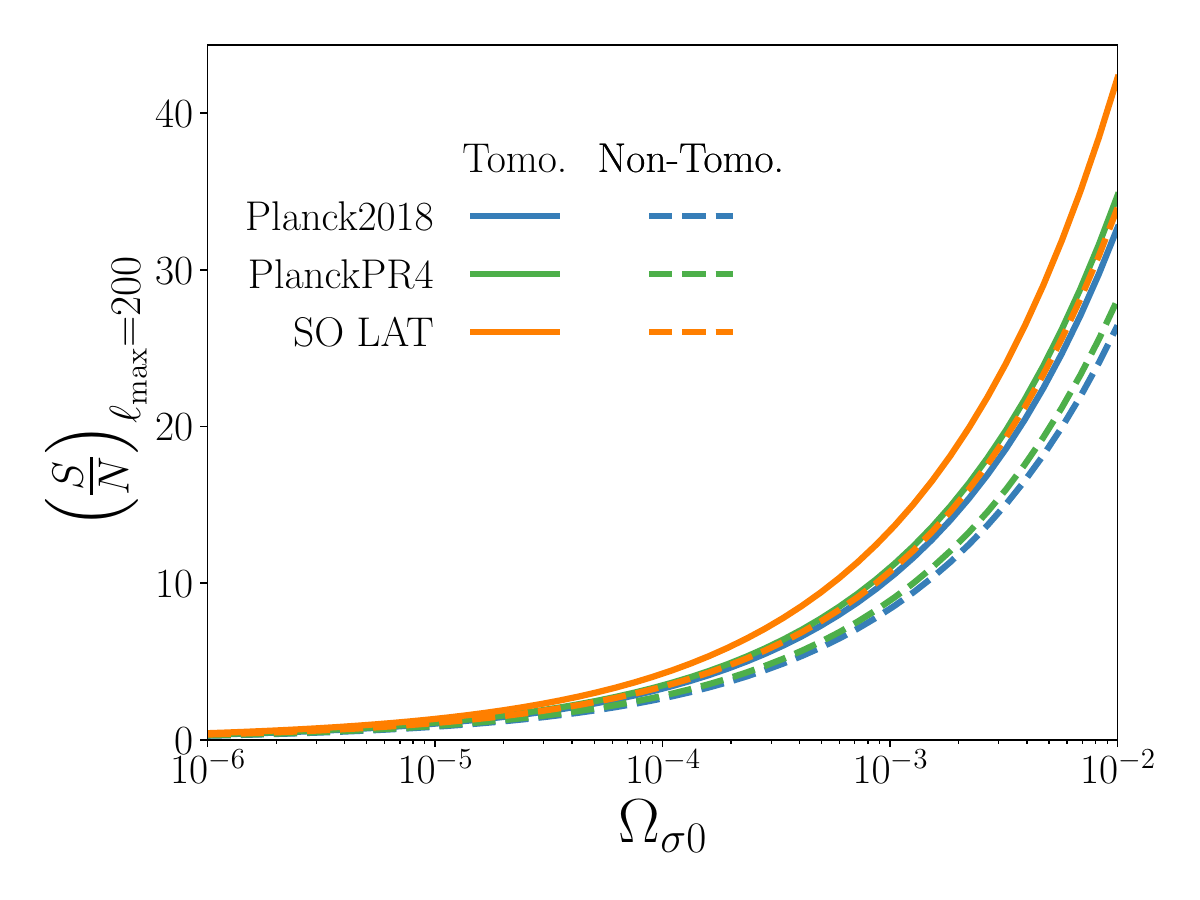}
    \caption{Tomographic (solid) and non-tomographic (dashed) cumulative $\hat{\P}^{\kappa i}_{\ell,0}$ SNRs summed from $\ell_{\text{min}}=10$ up to $\ell_{\text{max}} = 200$ as a function of $\Omega_{\sigma 0}$. The different colours denote different CMB surveys/datasets that have been cross-correlated with Euclid.}
    \label{fig:SNR_Tot_var_Omega_all_surveys}
\end{figure}

The cumulative SNRs, summed from $\ell_{\text{min}}=10$ to $\ell_{\text{max}} = 200$, are shown in \autoref{fig:SNR_Tot_var_Omega_all_surveys}. As expected, the predicted SNRs increase in tandem with the anisotropy parameter $\Omega_{\sigma 0}$. As one would predict from the CMB lensing reconstruction bias shown in \autoref{fig:Convergence_Spectra_and_Noise}, the Simons\,$\times$\,Euclid cross-correlation achieves a better SNR than other survey combinations when tomographic information is included. However, its performance is significantly hampered by the lower sky overlap between the Euclid and Simons survey footprints compared to Planck\,$\times$\,Euclid.

\begin{table}[htb!]
\centering
\begin{tabular}{@{}rlll@{}}
\toprule
                   & \multicolumn{1}{c}{\textbf{Planck2018}} & \multicolumn{1}{c}{\textbf{PlanckPR4}} & \multicolumn{1}{c}{\textbf{SO LAT}} \\ \midrule
\textbf{Non-Tomo.} & $3.8\%$                                  & $3.6\%$                                 & $2.9\%$                                  \\
\textbf{Tomo.}     & $3.1\%$                                  & $2.9\%$                                 & $2.4\%$                                  \\ \bottomrule
\end{tabular}\caption{Minimum detectable anisotropy ratios ${\sigma_{\perp0}}/{H_0}$ for threshold of $\text{SNR}=10$. }
\label{tbl:Min_sigma_H_SNR=10}
\end{table}

If we set a minimum signal-to-noise threshold of SNR$=10$, we can estimate the lowest detectable anisotropy value for each pair of surveys and evaluate the added value of tomographic information. These minimum anisotropy values are shown in \autoref{tbl:Min_sigma_H_SNR=10}. Note that the numerical values in this table correspond to the ratio of anisotropic to isotropic expansion today, ${\sigma_{\perp0}}/{H_0} = \sqrt{\Omega_{\sigma0}}\,$. When cross-correlated with Euclid, all CMB surveys appear to be sensitive at the percent level to the current anisotropy ratio. Including tomographic information enhances sensitivity by a factor of $\sim 1.2$ for all survey combinations --- a somewhat modest improvement.

This tentative forecast for the constraining power of this method comes \emph{solely} from the cross-correlation of the (non-tomographic) reconstructed CMB lensing convergence with the (tomographic) $B$-mode lensing signal measured from galaxy shapes. Including the $E$-$B$ cross-correlation signal obtained from galaxy ellipticity measurements, {as analysed in \cite{adam_probing_2025}}, would certainly increase the SNR and thereby improve constraining power on the anisotropy ratio, since inter-bin correlations significantly increase the number of combinations possible. However, in order to do this in a consistent manner, one would have to take into account the covariance of the $\kappa$-$B$ and $E$-$B$ estimators. We do not perform this full analysis here, but we outline the construction of the covariance matrix in \autoref{app:Covariance_Matrix}.

We note that $\hat{\P}^{\kappa i}_{\ell M}$, as defined in \autoref{eqn:P_lM_Estimator}, is the most basic statistical estimator for $\P^{\kappa i}_{\ell M}$ that can be constructed from the observed $\hat{\kappa}_{\ell m}$ and $\hat{B}^i_{\ell m}$ multipoles. More sophisticated estimators incorporating inverse-variance weighting and band-power averaging would improve the quality of the recovered signal, while a comprehensive analysis should also account for spurious anisotropies introduced by sky masking. 
Masking an anisotropic temperature signal was studied in \cite{aluri_novel_2015} and the effect on a dipolar anisotropy was found to be very mild. For the polarisation signal, a more general approach to masking, e.g. along the lines worked out in \cite{liu_methods_2019, liu_general_2019}, is in order. But these references also found that masks including $\gtrsim10\%$ of the sky lead only to a mild degradation, e.g. of the limit on the tensor to scalar ration $r$. 
Therefore
, we believe that our SNR forecasts demonstrate the viability of cross-correlating CMB lensing convergence with galaxy shear $B$-modes (together with the $E$-$B$ correlation) as a probe of late-time anisotropy.


\section{Conclusion}

In this work, we explored the potential of using cross-correlations between CMB lensing convergence and cosmic shear $B$-modes to test large-scale isotropy and, by extension, the Cosmological Principle. Although observations strongly favour isotropy in the early Universe, the possibility of late-time anisotropy, potentially arising from dark sector physics, remains a crucial area of investigation --- particularly in light of recent hints from kinematic dipole and bulk flow discrepancies. Our methodology models large-scale anisotropy using an axisymmetric Bianchi~I spacetime, enforcing rotational symmetry about a preferred axis and singling out a preferred direction. 

Building on our previous $E\times B$-mode analysis, we expanded the framework and offered qualitative insights into the resulting spectra. We showed that the anisotropy parameter $\Omega_{\sigma 0}$ ($=\sigma_{\perp0}^2/H_0^2$) acts as an overall scaling factor for $\mathcal{P}_{\ell M}^{\kappa i}$, preserving the spectral shape of the resulting SNR. Our forecasts show that the signal-to-noise ratio for $\hat{\mathcal{P}}_{\ell M}^{\kappa i}$ peaks around $\ell \sim 100$, a substantially larger scale than the standard $E$-mode lensing signal (which peaks around $\ell\sim 1100$). This offers a probe of anisotropic expansion distinct from the usual lensing power spectrum.

By defining a detectability threshold of $\text{SNR}\geq 10$, we were able to determine the minimum level of anisotropy which each survey combination is sensitive to. Among the configurations tested, the cross-correlation of the Euclid $B$-modes with the Simons convergence delivered the highest SNR despite limited sky overlap, achieving sensitivity to percent-level anisotropy. Making use of tomography improves detection power by a factor of $\sim 1.2$ for all survey combinations. This is without the inclusion of the $E$-$B$ cross-correlations --- adding them would further strengthen constraints.

While this study focused solely on the $\kappa$-$B$ cross-correlation, incorporating (tomographic) $E$-$B$ cross-correlation information provided by galaxy ellipticity measurements (as explored in previous work) would substantially increase the overall SNR and tighten constraints on anisotropy. A comprehensive future analysis would need to account for the covariance between the $\kappa$-$B$ and $E$-$B$ estimators in a consistent manner using the formalism outlined in \autoref{app:Covariance_Matrix}. Furthermore, making use of more sophisticated statistical estimators than those used here, along with accounting for the effects of a masked sky, promises to further improve the fidelity of the recovered signal.

Our results highlight the value of lensing-based observables as complementary probes of the cosmological principle. Future analyses combining multiple lensing sources and higher-quality data will provide even tighter tests of isotropy on the largest scales.

\acknowledgments 
We thank Julien Carron for providing us with the empirical estimate of the PlanckPR4 lensing convergence bias.
JA and RM are supported by the South African Radio Astronomy Observatory and National Research Foundation (grant no. 75415).
Numerical calculations were performed using \texttt{NumPy} \cite{harris_array_2020} and \texttt{SciPy} \cite{virtanen_scipy_2020}. Figures were produced using \texttt{matplotlib} \cite{hunter_matplotlib_2007}.


\newpage
\cleardoublepage
\appendix

\section{\boldmath Full covariance matrix for \texorpdfstring{$\kappa$}{kappa}-\texorpdfstring{$B$}{B} and \texorpdfstring{$E$}{E}-\texorpdfstring{$B$}{B} estimators}\label{app:Covariance_Matrix}

Suppose we have estimated $\hat{\mathcal{P}}_{\ell M}^{\kappa i}$ and $\hat{\mathcal{P}}_{\ell M}^{i j}$ from $\kappa$-$B$ and $E$-$B$ cross-correlations, respectively. We can collect all of these quantities into the data vector $\hat{\bm{d}}$ by stacking them into a column
\begin{equation}
    \hat{\bm{d}} = \mqty( \big[\hat{\mathcal{P}}_{\ell M}^{\kappa i} \big]  \\\big[\hat{\mathcal{P}}_{\ell' M'}^{j k}\big]).
\end{equation}
The covariance matrix associated with this data vector is then 
\begin{equation}
    \text{Cov}\big(\hat{\bm{d}} , \hat{\bm{d}}^{\dagger}\big) = \expval*{\hat{\bm{d}}\,\hat{\bm{d}}^{\dagger}} - \langle{\hat{\bm{d}}}\rangle\langle{\hat{\bm{d}}^{\dagger}}\rangle,
\end{equation}
where the superscript `$\dagger$' denotes the Hermitian conjugate. The statistically-isotropic part of this covariance matrix is given by
\begin{equation}
    \text{Cov}\big(\hat{\bm{d}} , \hat{\bm{d}}^{\dagger}\big)_{\text{SI}} = \mqty(\text{Cov}\big( \hat{\mathcal{P}}_{\ell M}^{\kappa i}, \hat{\mathcal{P}}_{\ell' M'}^{\kappa j}\big)_{\text{SI}} & \text{Cov}\big( \hat{\mathcal{P}}_{\ell M}^{\kappa i}, \hat{\mathcal{P}}_{\ell' M'}^{j k}\big)_{\text{SI}}\\ & \\
   \text{Cov}\big( \hat{\mathcal{P}}_{\ell M}^{i l}, \hat{\mathcal{P}}_{\ell' M'}^{\kappa j}\big)_{\text{SI}} & \text{Cov}\big( \hat{\mathcal{P}}_{\ell M}^{il}, \hat{\mathcal{P}}_{\ell' M'}^{jk}\big)_{\text{SI}}),
\end{equation}
where
\begin{subequations}
    \begin{align}
\text{Cov}\big(\hat{\mathcal{P}}_{\ell M}^{\kappa i}, \hat{\mathcal{P}}_{\ell' M'}^{\kappa j}\big)_{\text{SI}} &= \frac{20}{f_{\text{sky}}} \left[ \frac{\ell^2 (\ell+1)^2(\ell+2)!}{(\ell-2)!} \right]^{-1} \largesum_{I=\pm 1}\frac{{\left(C^{\kappa \kappa}_{\ell}\right)_{\text{SI}}} {\big(C^{B^i B^j}_{\ell+I}\big)_{\text{SI}}}}{\left({^2F_{\ell+I, 2 ,\ell}}\right)^2}\, \delta_{\ell \ell'}\delta_{MM'},\\
        \text{Cov}\big(\hat{\mathcal{P}}_{\ell M}^{ij}, \hat{\mathcal{P}}_{\ell' M'}^{kl}\big)_{\text{SI}} &= \frac{20}{f_{\text{sky}}} \left[ \frac{(\ell-2)!}{(\ell+2)!} \right]^2 \largesum_{I=\pm 1}\frac{{\big(C^{E^i E^k}_{\ell}\big)_{\text{SI}}} {\big(C^{B^j B^l}_{\ell+I}\big)_{\text{SI}}}}{\left({^2F_{\ell+I, 2 ,\ell}}\right)^2}\, \delta_{\ell \ell'}\delta_{MM'},\\
        \text{Cov}\big(\hat{\mathcal{P}}_{\ell M}^{\kappa i}, \hat{\mathcal{P}}_{\ell' M'}^{jk}\big)_{\text{SI}} &=  -\frac{20}{f_{\text{sky}}} \frac{1}{\ell(\ell+1)}\left[\frac{(\ell-2)!}{(\ell+2)!} \right]^{3/2} \largesum_{I=\pm 1}\frac{{\big(C^{\kappa E^j}_{\ell}\big)_{\text{SI}}} {\big(C^{B^i B^k}_{\ell+I}\big)_{\text{SI}}}}{\left({^2F_{\ell+I, 2 ,\ell}}\right)^2} \delta_{\ell \ell'}\delta_{MM'},
    \end{align}
\end{subequations}
and
\begin{subequations}
    \begin{align}
        \expval{\hat{\kappa}_{\ell m} \hat{\kappa}^{*}_{\ell' m'}}_{\text{SI}} &= {\left(C^{\kappa\kappa}_{\ell}\right)_{\text{SI}}}\delta_{\ell \ell'}\delta_{m m'} = \left(C^{\kappa\kappa}_\ell + N^{\kappa\kappa}_\ell\right)\delta_{\ell \ell'}\delta_{m m'}\,, \\
        \expval{\hat{E}^i_{\ell m} \hat{E}^{j*}_{\ell' m'}}_{\text{SI}} &= {\left(C^{E^iE^j}_{\ell}\right)_{\text{SI}}}\delta_{\ell \ell'}\delta_{m m'} = \left(C^{E^iE^j}_\ell + \frac{\expval{\gamma_{\text{int.}}^2}}{\bar{N}_{i}}\delta_{ij}\right)\delta_{\ell \ell'}\delta_{m m'}\,, \label{eqn:E_Mode_Variance}\\
        \expval{\hat{\kappa}_{\ell m} \hat{E}^{i*}_{\ell' m'}}_{\text{SI}} &= {\left(C^{\kappa E^i}_{\ell}\right)_{\text{SI}}}\delta_{\ell \ell'}\delta_{m m'} = C^{\kappa E^i}_\ell\delta_{\ell \ell'}\delta_{m m'}\,, \\
        \expval{\hat{B}^i_{\ell m} \hat{B}^{j*}_{\ell' m'}}_{\text{SI}} &= {\left(C^{B^i B^j}_{\ell}\right)_{\text{SI}}}\delta_{\ell \ell'}\delta_{m m'} = \frac{\expval{\gamma_{\text{int.}}^2}}{\bar{N}_{i}}\delta_{ij}\delta_{\ell \ell'}\delta_{m m'}\,, \\        
        \expval{\hat{\kappa}_{\ell m} \hat{B}^{i*}_{\ell' m'}}_{\text{SI}} &= {\left(C^{\kappa B^i}_{\ell}\right)_{\text{SI}}}\delta_{\ell \ell'}\delta_{m m'} = 0\,,\\
        \expval{\hat{E}^i_{\ell m} \hat{B}^{j*}_{\ell' m'}}_{\text{SI}} &= {\left(C^{E^i B^j}_{\ell}\right)_{\text{SI}}}\delta_{\ell \ell'}\delta_{m m'} = 0\,.
    \end{align}
\end{subequations}
Under the (reasonable) assumption that the systematic errors/biases of the convergence and $E$-modes are uncorrelated, their statistically-isotropic power spectrum is given by
\begin{equation}
    {\left(C^{\kappa E^i}_{\ell}\right)_{\text{SI}}}  = \int \dd k\, k^2 P(k) \Delta_\ell^\kappa(k) \Delta_\ell^i(k),
\end{equation}
where we have made use of the $E$-mode tomographic kernel
\begin{equation}
    \Delta^{i}_\ell(k) = \frac{1}{2}\bigg[{\frac{2}{\pi}}\,{\frac{(\ell+2)!}{(\ell-2)!}}\bigg]^{1/2}\int_{0}^{\chi_s}\dd\chi \, q^i(\chi,\chi_s)\,j_{\ell}(k\chi) \, T_\varphi(\chi,k).
\end{equation}


\section{Additional figures}

\begin{figure}[htb!]
	\centering
	\includegraphics[width=0.6\linewidth]{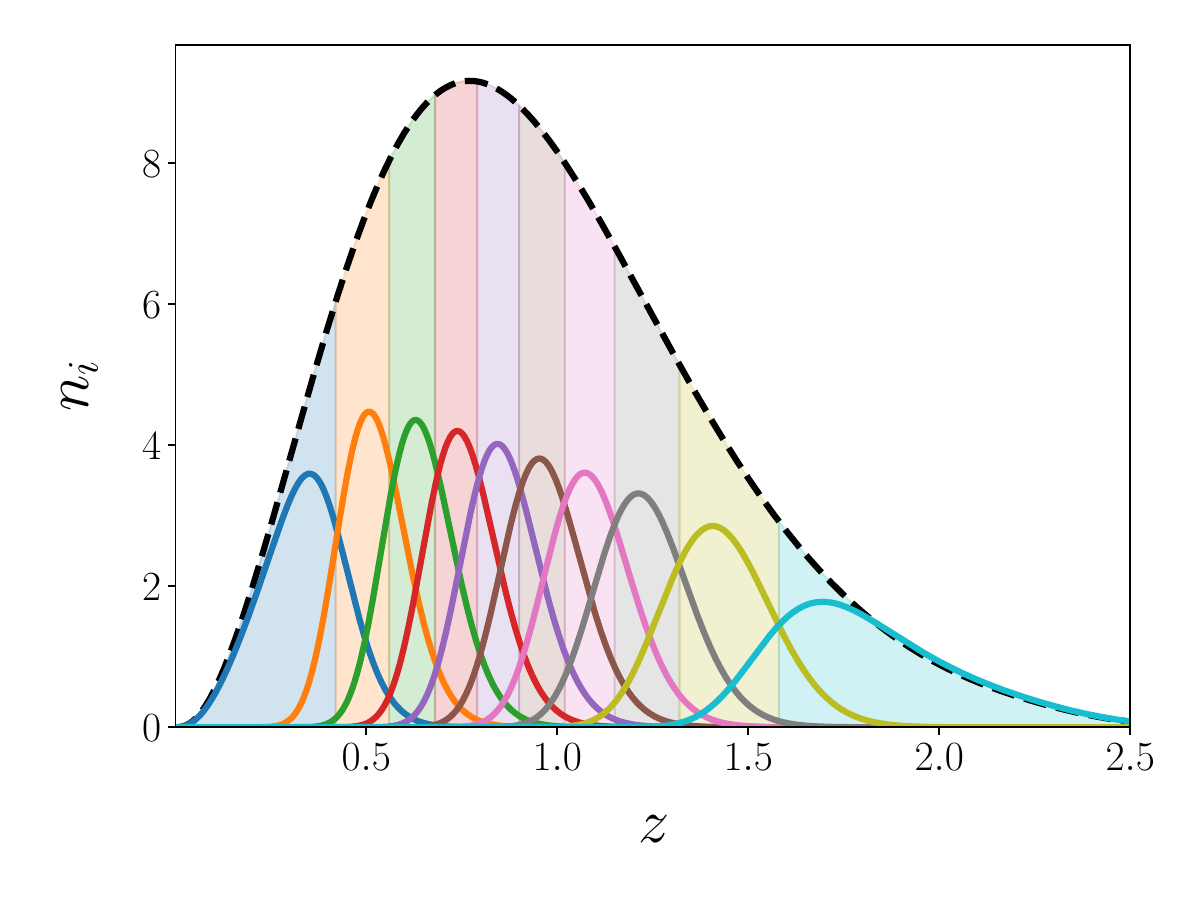}
	\caption{Decomposition of the full source redshift distribution $n_\text{F}$ (dashed line) into 10 equi-populated tomographic bins $n_i$ (coloured lines). For visualisation, $n_\text{F}$ has been scaled by a factor of 10 so that its area matches the combined areas of the tomographic distributions. Shaded regions indicate the bin ranges, with each coloured curve encompassing the same area as its corresponding shaded region.}
 \label{fig:Euclid_Tomographic_Distributions}
\end{figure}

\begin{figure}[htb!]
    \centering
    \includegraphics[width=\linewidth]{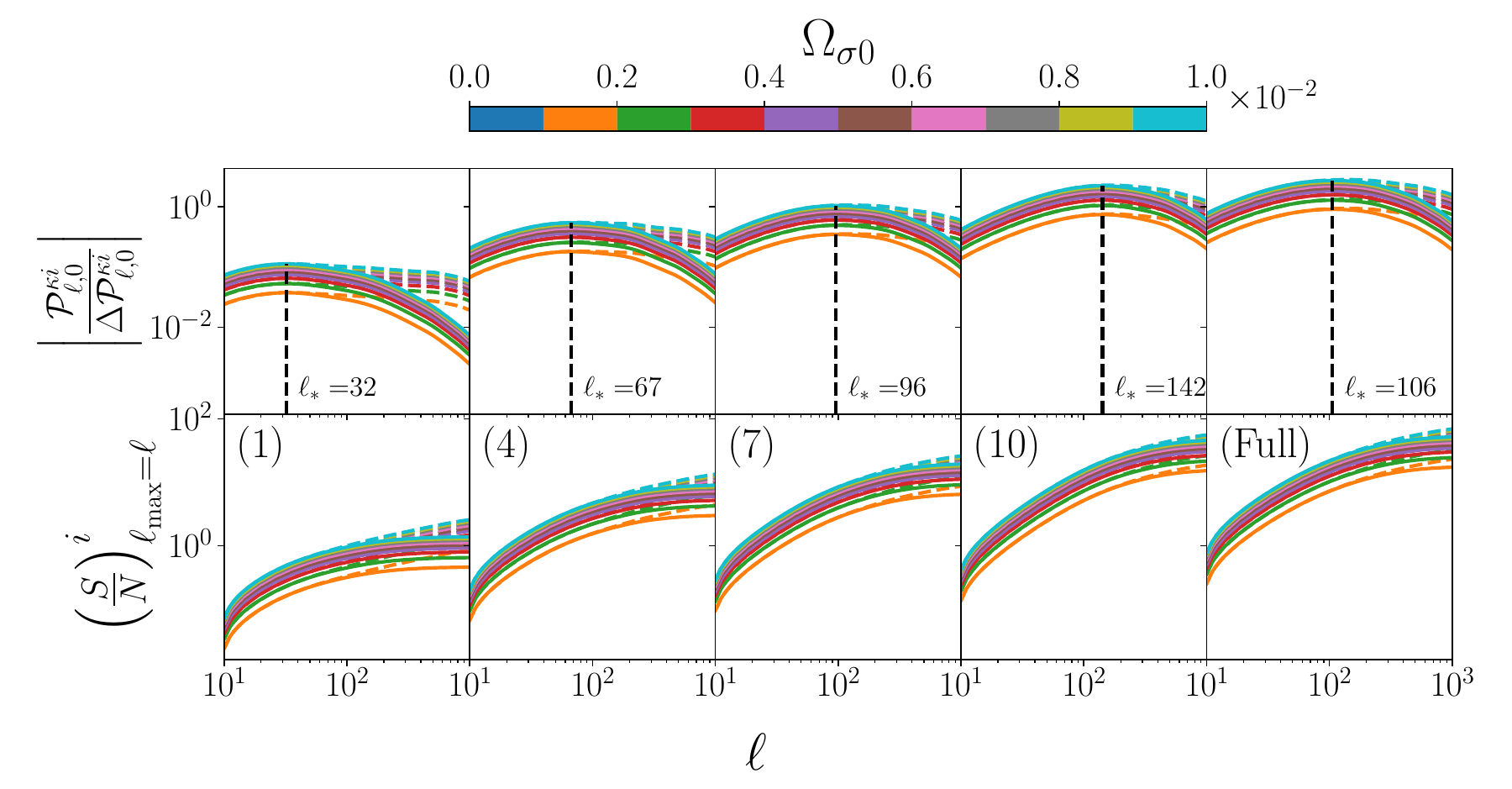}
    \caption{Individual (top) and cumulative (bottom) Simons\,$\times$\,Euclid signal-to-noise ratios for $\hat{\P}^{\kappa i}_{\ell,0}$ computed for redshift bins $i=1, 4, 7, 10$, as well as the full redshift range. The linear (solid) and HaloFit (dashed) ratios have been calculated for ten values of $\Omega_{\sigma 0}\leq 10^{-2}$. Note that we use $\ell_{\text{min}}=10$ for the cumulative SNR.}
    \label{fig:SNR_Tomo_Full}
\end{figure}

\begin{figure}[htb!]
    \centering
    \includegraphics[width=\WidthScale\linewidth]{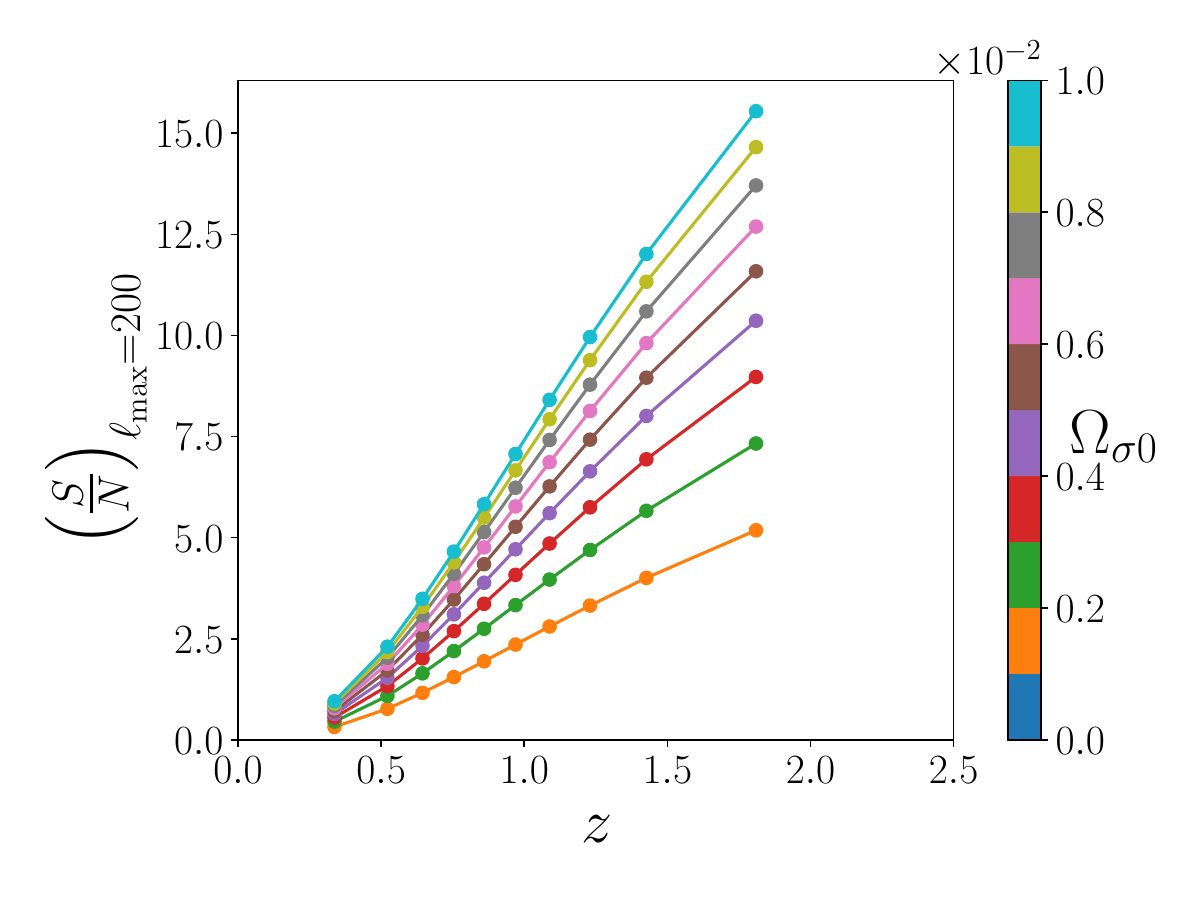}
    \caption{Cumulative SNRs of $\hat{\P}^{\kappa i}_{\ell,0}$ for Simons\,$\times$\,Euclid, summed over multipoles from $\ell_{\text{min}}=10$ to $\ell_{\text{max}}=200$, for the 10 Euclid redshift bins and 10 values of $\Omega_{\sigma 0}\leq 10^{-2}$. Each point is shown at the mean redshift of its corresponding bin.}
    \label{fig:SNR_Var_Bins}
\end{figure}


\begin{figure}[htb!]
    \centering
    \includegraphics[width=0.6\linewidth]{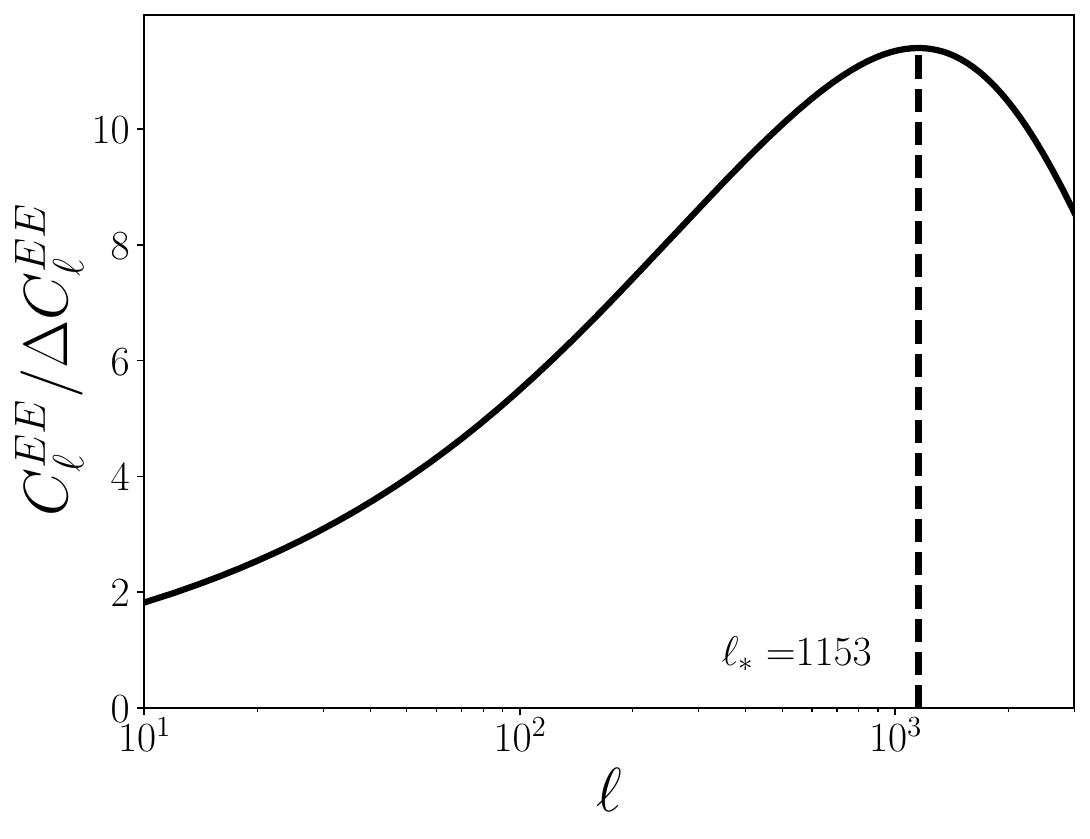}
    \caption{Signal-to-noise ratio for the standard $E$-mode lensing power spectrum $C^{EE}_\ell$, computed for the full (non-tomographic) Euclid source distribution, as a function of multipole moment $\ell$. The noise $\Delta C_\ell^{EE}$ is calculated by dividing \autoref{eqn:E_Mode_Variance} by the factor $f_{\text{sky}}(2\ell+1)$ and taking the square root. Nonlinear effects are included via HaloFit. The multipole $\ell_*$ corresponding to the maximum individual SNR is marked with a vertical dashed line. The \texttt{CLASS} precision parameter \texttt{k\_max\_tau0\_over\_l\_max} is set to 10.0 to properly capture small-scale power.}
    \label{fig:C^EE_l_SNR}
\end{figure}

\begin{figure}[htb!]
    \centering
    \includegraphics[width=0.6\linewidth]{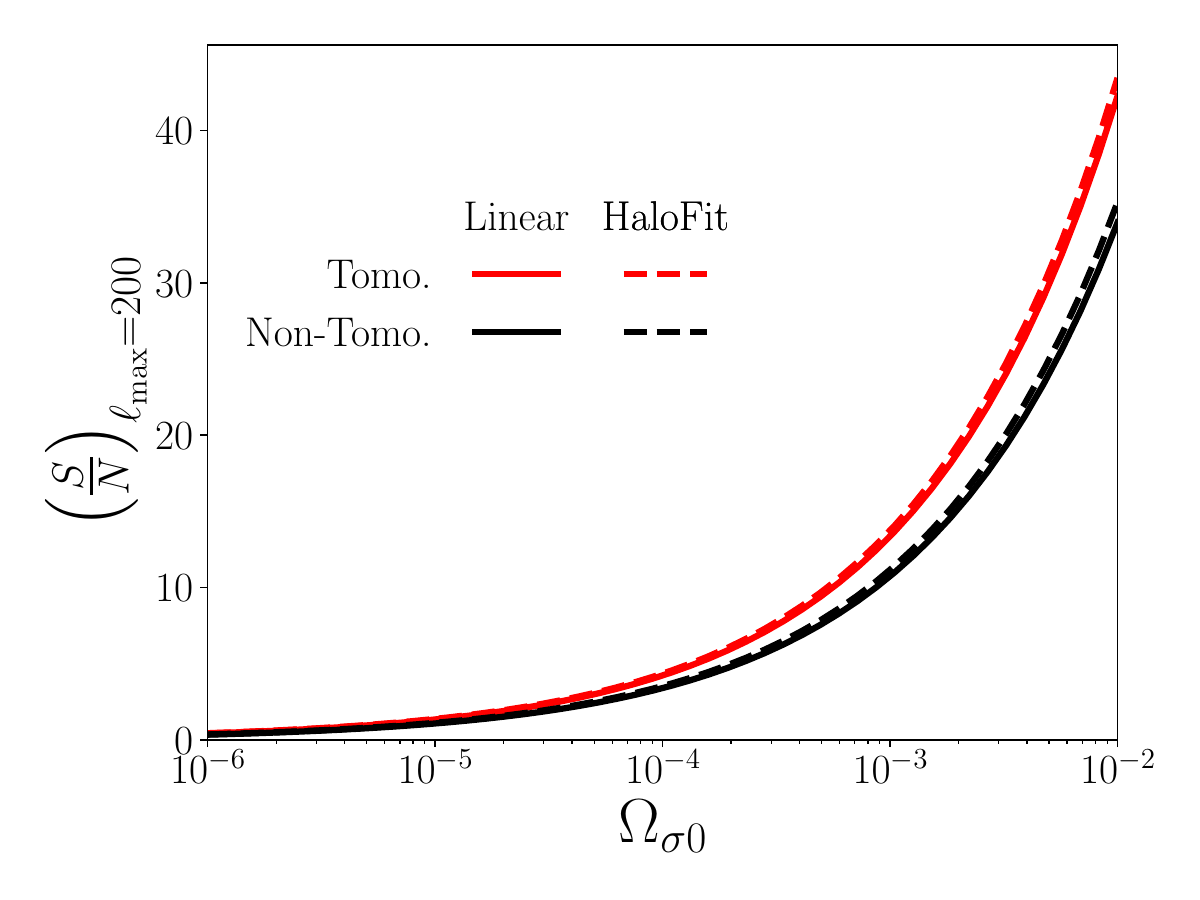}
    \caption{Linear (solid) and HaloFit (dashed) cumulative $\hat{\P}^{\kappa i}_{\ell,0}$ signal-to-noise ratios summed from $\ell_{\text{min}}=10$ up to $\ell_{\text{max}} = 200$ as a function of $\Omega_{\sigma 0}$ calculated using the SO LAT noise curve. The red lines show the SNR obtained by summing up all tomographic bins while the full (non-tomographic) SNR is shown in black.}
    \label{fig:SNR_Tot_var_Omega_SO}
\end{figure}

\clearpage
\bibliographystyle{JHEP}
\bibliography{kappaxB_references}

\end{document}